\documentclass[conference]{IEEEtran}
\IEEEoverridecommandlockouts
\usepackage[utf8]{inputenc}
\usepackage{mathcomSTEv4}

\usepackage{caption}
\usepackage{subcaption}
\usepackage{verbatim}
\usepackage{todonotes}
\usepackage{multirow}
\usepackage{multicol}
\usepackage{color}

\usepackage{graphics}
\usepackage{amsmath,amsfonts, amssymb}
\usepackage{mathtools}
\usepackage{epstopdf}
\usepackage{amsthm}
\usepackage{bbm}
\usepackage{algorithm}
\usepackage[noend]{algpseudocode}
\usepackage{dsfont}
\usepackage{cite}
\usepackage{subcaption}

\newcommand{\thehv}{\widehat{\thev}}

\def\BibTeX{{\rm B\kern-.05em{\sc i\kern-.025em b}\kern-.08em
		T\kern-.1667em\lower.7ex\hbox{E}\kern-.125emX}}
\begin{document}
	
	\title{Decentralized SGD with Over-the-Air Computation
		\thanks{This work has been funded by the European Research Council (ERC) through project BEACON (No. 725731) and by by H2020-MSCA-ITN-2015 project SCAVENGE under grant number 677854.}
		\author{\IEEEauthorblockN{ Mehmet Emre Ozfatura,$^+$ Stefano Rini$^\dagger$,  Deniz G\"{u}nd\"{u}z$^+$}
			\IEEEauthorblockA{\textit{$^+$ Department of Electrical and Electronic Engineering, Imperial College London, UK}\\
				\textit{$^\dagger$ Department of Electrical and Electronic Engineering, NCTU, Taiwan}
	}}}
	
	\maketitle
		
	\begin{abstract}
		We study the performance of decentralized stochastic gradient descent (DSGD) in a wireless network, where the nodes collaboratively optimize an objective function using their local datasets. Unlike the conventional setting, where the nodes communicate over error-free orthogonal communication links, we assume that transmissions are prone to additive noise and interference. We first consider a point-to-point (P2P) transmission strategy, termed the OAC-P2P scheme, in which the node pairs are scheduled in an orthogonal fashion to minimize interference. Since in the DSGD framework, each node requires  a linear combination of the neighboring models at the consensus step, we then propose the OAC-MAC scheme, which utilizes the signal superposition property of the wireless medium to achieve over-the-air computation (OAC). For both schemes, we cast the scheduling problem as a graph coloring problem. 
		We numerically evaluate the performance of these two schemes for the MNIST image classification task under various network conditions.
		We show that the OAC-MAC scheme attains better convergence performance with a fewer communication rounds.
	\end{abstract}
	
	\section{Introduction}
	Over the last decade, modern machine learning (ML) techniques, particularly deep learning (DL) framework, have demonstrated tremendous advances in many complex problems such as machine vision, speech recognition, and natural language processing. From the computational aspect, the main task in DL framework is computing the gradients through the layers in the back-propagation step, which is later used to update the network parameters. Since it is practically not possible to train the network over a huge data set, stochastic gradient descent (SGD) is used instead, in which a random subset of data, called a mini-batch, is used to compute an estimate of the gradient at each step. To further reduce the training time, stochastic gradient computations can be distributed to multiple \textit{workers}, each of which executes the back-propagation step over a different subset of data, in parallel, under orchestration of the parameter server that is responsible for the model update \cite{PSGD2}. This approach is often referred to as parallel SGD (PSGD) in the literature. PSGD is also employed when data is already distributed at multiple devices, which would like to train a model collaboratively without offloading their data to a central server due to privacy concerns. This latter framework is known as \textit{federated learning} \cite{KonecnyFLBeyondData, DCKonecnyFederated}.
 
Although the PSGD framework can successfully reduce the computation time and address the privacy concerns, it requires large amount of data transfer between the parameter server and the workers at each iteration, thus the communication latency becomes the new bottleneck. Hence, the recent studies have focused on the communication latency of PSGD, and introduced different strategies to mitigate it, such as  sparsification \cite{SGD_sparse1}, where gradients are transformed into sparse vectors by keeping only the significant values, and quantization \cite{SGD_q1}, where each gradient is represented by only a limited number of bits. 
 
While federated learning has been extremely popular over the last few years, the conventional parameter server architecture, in which a central server orchestrates the training across all the participating nodes has its limitations. First of all, such a parameter server may not always be available. Moreover, the centralized structure creates congestion at the parameter server, and leads to a single-point-of-failure, limiting the robustness of the system, and increasing the security risks. 
 
An alternative framework is decentralized SGD (DSGD), where each worker communicates with only a certain number of other workers without assistance of a parameter server \cite{decent3, decent7, decent12, decent14}. Such a decentralized architecture can address the congestion at the parameter server, and reduce the communication latency by relaxing the requirement to communicate with a single node from every other collaborating node in the network.\\ 
\indent Although there is a broadening literature on communication efficient DSGD framework, in none of these papers wireless nature of the communication medium is taken into account. However, when learning takes place at the wireless network edge \cite{ML_edge}, it is indispensable to take into account the wireless nature of the communication medium, which creates additional challenges due to the inherent noise and interference among nodes. In this paper, we study DSGD over wireless networks taking into account the physical channel characteristics into account, and propose communication efficient transmission and scheduling schemes to reduce the communication delay of the learning algorithm.



	\subsubsection*{Contributions}
	In this paper, we first highlight how the communication strategy between the nodes affects the convergence performance of the DSGD scheme when implemented in a wireless network. 
	%
	%
	%
	We first introduce a point-to-point(P2P) strategy, in which pairs of nodes are scheduled with the goal of avoiding or minimizing interference.  
	We then propose an alternative multiple access (MAC) strategy, where multiple nodes transmit simultaneously to a common receiver, such that their transmissions are aligned for over-the-air computation (OAC). This MAC strategy harnesses interference from other nodes, rather than trying to mitigate it. It has previously been used in the federated learning setting in \cite{ML_edge, MohammadDenizDSGDCS,zhu2019broadband}. 
	Note that the MAC strategy improves the efficiency of resource usage by scheduling all the transmitters at a single time slot, and provides better noise mitigation. Such simultaneous transmission does not lead to any additional interference since the receiver is not interested in the individual estimates of the other nodes, but only in their weighted average. We will achieve different weights as required by the consensus algorithm by allowing scaling at the transmitters. Note that, while multiple receivers can be scheduled simultaneously in the P2P case, this is more challenging in the MAC case.
	
	We evaluate the performance of these two strategies for the MNIST digit classification task under various network conditions. We observe that the MAC approach with OAC provides significant gains in terms of the wall-clock convergence speed. It also results in higher accuracy (lower test error) due to its increased robustness against channel noise.

	\section{Problem Formulation}
	\label{sec:Problem formulation}
	
	\subsection{System Model}
	\label{sec:System Model}
	We consider the following distributed stochastic optimization problem 
	\begin{equation}
   \min_{\theta\in\mathbb{R}^{d}} f(\thev) \triangleq \frac{1}{n}\sum^{n}_{i=1}\underbrace{\mathds{E}_{\zeta \sim \mathcal{D}_{i}}F(\thev,\zeta)}_{\mathrel{\mathop:}=f_{i}(\thev)},\label{eq:global function}
    \end{equation}
	where $f_{i}(\theta)$ is the loss function at node $i\in[n] \triangleq \{1, \ldots, n\}$ based on the local sample distribution $\mathcal{D}_{i}$, and $F(\thev,\zeta)$ is parameterized loss function imposed by the learning task and $\zeta$ is variable referring the data sample. When the samples are from a finite data set, as in many DL problem, $f_{i}(\thev)= \frac{1}{\vert D_{i} \vert} \sum_{\zeta\in D_{i}}F(\thev,\zeta)$. We assume that $f_i(\theta)$ is convex and $L$-smooth, $\forall i\in[n]$. We  also assume that the solution of \eqref{eq:global function} is finite, and denoted by $\thev^*$. The goal of each node is to obtain an estimate of $\thev^*$. In particular, we wish to minimize the maximum disagreement defined as
	\ea{
		\ep(t)=	\max_i \| f \lb \thehv_i (t)\rb-f \lb \thev^* \rb \| ,
	}
	where $\thehv_i (t) \triangleq 1/t \sum_{l \in [t]} \thev_i (l) $, and
	$\thev_i(t)$ is the estimate of $\thev^*$ at node $i$ at time $t$.\\
	\indent We implement DSGD in three steps to solve \eqref{eq:global function} in a distributed manner. First, at the beginning of each iteration, node $i$ computes the {\em local stochastic gradient}, i.e.,
	\begin{equation}
	g_i(t) \triangleq \nabla_{\theta}F(\thev_{i}(t),\zeta_{i,t}),
	\label{eq:stochastic gradient}
	\end{equation}
	where $\zeta_{i,\tau}$ is randomly  sampled data at iteration $t$ by the $i$th worker. And it is often assumed that stochastic gradient in \eqref{eq:stochastic gradient} have a bounded variance.
Then, in the intermediate step, each node exchanges its local model with the other nodes to seek a consensus, i.e.,
\begin{equation}
\thev_{i}(t+1/2)=\sum_{j\in[n]} m_{i,j} \thev_{j}(t),\label{consensus}
\end{equation}
and, finally updates the local model, i.e., $\thev_{i}(t+1)=\thev_{i}(t+1/2)+g_{i}(t)$. We note that the coefficients $m_{i,j}$ depend on the connectivity pattern between the nodes; specifically, if the network is defined by a connected graph $\Gcal=(V,E)$ with $V = \{V_1, \ldots, V_n\}$ and $E \subseteq  V \times V$, the coefficients are chosen according to the adjacency matrix $\Av$ of $\Gcal$, so that they form a {\em doubly stochastic matrix}, which is the necessary condition for the convergence of DSGD.

	%
\subsection{DSGD over a Wireless Network}

Here, we will briefly  explain how the wireless communication model and DSGD framework are linked to each other. Without any topology constraint on the network, the nodes are allowed to communicate with each other over a fading additive white Gaussian noise (AWGN) channel. The communication across nodes is described as follows: 
	%
	at time $t \in \Nbb$, node $V_i$, $i \in [n]$, transmits an input signal $\xv_{i}(t) \in \Rbb^d$, and receives the output signal $\yv_{i}(t) \in \Rbb^d$ obtained as
	\ea{
		\yv_{i}(t) = \sum_{j \in V \setminus \{V_i\}} h_{ij}  \xv_{j}(t)+\nv_{i}(t), 
		\label{eq:in/out}
	}
	where  $h_{ij}=h_{ji}$ is the channel gain between nodes $i$ and $j$, and $\nv_{j}(t) \sim \Ncal(\zerov_d,\Iv_{s}^d)$ represents the additive noise vector. We assume that the channel gains $h_{ij}$ are independent and identically distributed (i.i.d.) according to distribution $P_h$, and remain fixed until the convergence is attained. We further assume that each node knows the coefficients of its incoming/ outgoing channels. We consider full-duplex nodes; that is, they can simultaneously transmit and receive signals.
	
		The channel input at each node $i$ is subject to the per-symbol power constraint
	\ea{
		\Ebb \lsb  \| \xv_{i}(t) \|^2 \rsb \leq P, \quad \forall \ t  \in \Rbb, \ i \in  [n],
		\label{eq:max power}
	}
	where $\| \cdot \|$  is the $2$-norm.
	Note that the expected value in \eqref{eq:max power} is with respect  to the additive noise and the stochastic gradient.

	Assume that the communication phase for the consensus step given in (\ref{consensus}) lasts $T$ time instants per computation round, referred to as a communication block. $T$ denotes the resources required for the communication phase, and has a direct impact on the wall-clock convergence speed. At time $t=(m+1)T$, node $i$ possesses the $m$ channel outputs obtained in the previous communication block $\Yv_i(mT)=[\yv_i(mT+1), \ldots , \yv_i((m+1)T)],$
	where
	\ea{
		\yv_i(mT+k) & =\sum_{j\in  V \setminus \{V_i\}} h_{ij} \xv_j(mT+k)+\nv_i(mT+k), \nonumber \\
		& =\sum_{j\in  V \setminus \{V_i\}} h_{ij} w_{jk} \thev_j(m)+\nv_i(mT+k),
		\label{eq:output channel block}
	}
\noindent
where $\thev_j(m)$ is the local estimate of $\thev^*$ at $V_j$ at communication round $m$, and $w_{jk}$ is the coefficient with which user $j$ scales its estimate for the $k$th symbol of the $m^{\rm th}$ communication round. Using $\Yv_i(mT)$, node $V_i$ updates its estimate as 
\ea{\small	
		 \thev_i(m+1) &= p_{ii} \thev_i(m) + \sum_{k \in [T]} p_{ik} \yv_i(mT+k) +\al_i(m)g_i(t), 
		\label{eq:estimate update}
	}
where $P \in \Rbb^{d \times m}$ are the combining weights, while $\al_i(m)$ is the learning rate in block $m$, and $g_i(t)$ is the stochastic gradient  in \eqref{eq:stochastic gradient}.
	The update in \eqref{eq:estimate update} can be rewritten as 
	\ea{
		& \thev_i(m+1) 
		= \beta_i \thev_i(m) + \sum_{j\in V \setminus \{i\}, k \in [T]} p_{ik} h_{ij} w_{jk} \thev_j(m) \nonumber \\
		& \quad \quad  \quad \quad + \sum_{k \in [T]} p_{ik} \nv_{ik}(mT+k),
		+\al_i(m)g_i(m) \label{eq:DOGD equivalent}  \\
		& = p_{ii} \thev_i(m) + \sum_{j\in V \setminus \{i\}} \pt_{ij}  \thev_j(m)+ \ntv_i(m) +\al_i(m)g_i(m),		\nonumber
	}
	where
	\ea{
	\pt_{ik} \triangleq \sum_{k \in [T]}p_{ik} h_{ij} w_{jk}, \quad \pt_{ii} \triangleq p_{ii},
}
	 If the matrix $\Ptv \in \Rbb^{d \times d}$ with entries $\pt_{ik}$ is doubly stochastic, then convergence for DSGD can be shown \cite{decent14}.
	We note that, in \eqref{eq:DOGD equivalent},  $\ntv_{i}(m)	= \sum_{k \in [T]} p_{ik} \nv_{ik}(mT +K)$. In the communication phase, by choosing the elements of matrix $\Wv \in \Rbb^{n \times T}$, denoted by $w_{jk}$, $j\in[n]$, and the elements of $\Pv$, denoted by $p_{ik}$, $k\in[T]$, a different weight matrix $\Ptv$ and a different noise matrix $\Ntv\in\Rbb^{n \times d}$ in \eqref{eq:DOGD equivalent} are obtained. Convergence of DSGD in the wireless setting can be analyzed using the existing techniques for the convergence of  DSGD framework \cite{decent14}, which we omit here due to space limitation. Nevertheless, regarding the communication phase of the proposed approach, the fundamental design question is \emph{how to choose $\Wv$, $\Pv$, and $T$ to achieve the fastest convergence possible, while guaranteeing that $\Ptv$ is a doubly stochastic matrix.}
	
	\subsection{Problem Definition}
\indent In the scope of this paper, we assume that, based on the channel realization $\Hv$, certain links between the nodes will not be used for communications, particularly those having channel gains below a predetermined threshold $h_{th}$. These links will only act as interference, and their impact will be included in the noise term.

Under this assumption, network topology can be considered as a random graph $\Gcal$ with  adjacency matrix $\Av$, and we can decouple the problem into two sub-problems. Accordingly, we first construct a symmetric doubly-stochastic $\Ptv$ from the matrix Laplacian as
\ea{
\Ptv=\Iv-\frac{1}{d_{\max}+1}(\Dv-\Av),
\label{eq:laplacian}
}
where $\Dv$ is a diagonal matrix whose entry in position $D_{ii}$ is the degree of $V_i$, and $d_{\max}$ is the maximum node degree.

Once $\Ptv$ is fixed, we can search for an efficient communication strategy, in terms of $T$ and $\Ntv$, satisfying predetermined weight matrix $\Ptv$. We particularly focus on drawing a connection between the duration of the communication block $T$ and the graph coloring problem. From a high-level perspective, communication strategies answering this question can be classified as follows:

		
		\noindent
		$\bullet$ {\bf P2P scheme:} 
		Only pair-wise transmissions are allowed. At each instant, two pairs of users transmit simultaneously only if their transmitters do not interfere. Here we consider as interfering only those links with gains above $h_{th}$. 

		\noindent
		$\bullet$
		{\bf MAC-OAC scheme:} Multiple nodes concurrently transmit their estimates in an uncoded `analog' manner, and the receiver receives a linear combination of the transmitted model (together with some noise) utilizing the superposition property of the  wireless  channel.
		Two nodes can receive simultaneously only if none of their transmitters interfere.
		
		\noindent
		$\bullet$
		{\bf Broadcast (BC)-OAC scheme:} A transmitter sends its signal to multiple receivers.
		Two nodes can broadcast simultaneously if they do not interfere at any of their receivers.
		
	\noindent
	$\bullet$	
		{\bf Interference Channel (IFC)-OAC scheme:} Two nodes transmit or receive simultaneously even if they interfere. 

	Due to space limitations, we will consider only P2P and MAC communication schemes in this paper. 
	
	\begin{figure}[]
		\centering
		\includegraphics[scale=0.35]{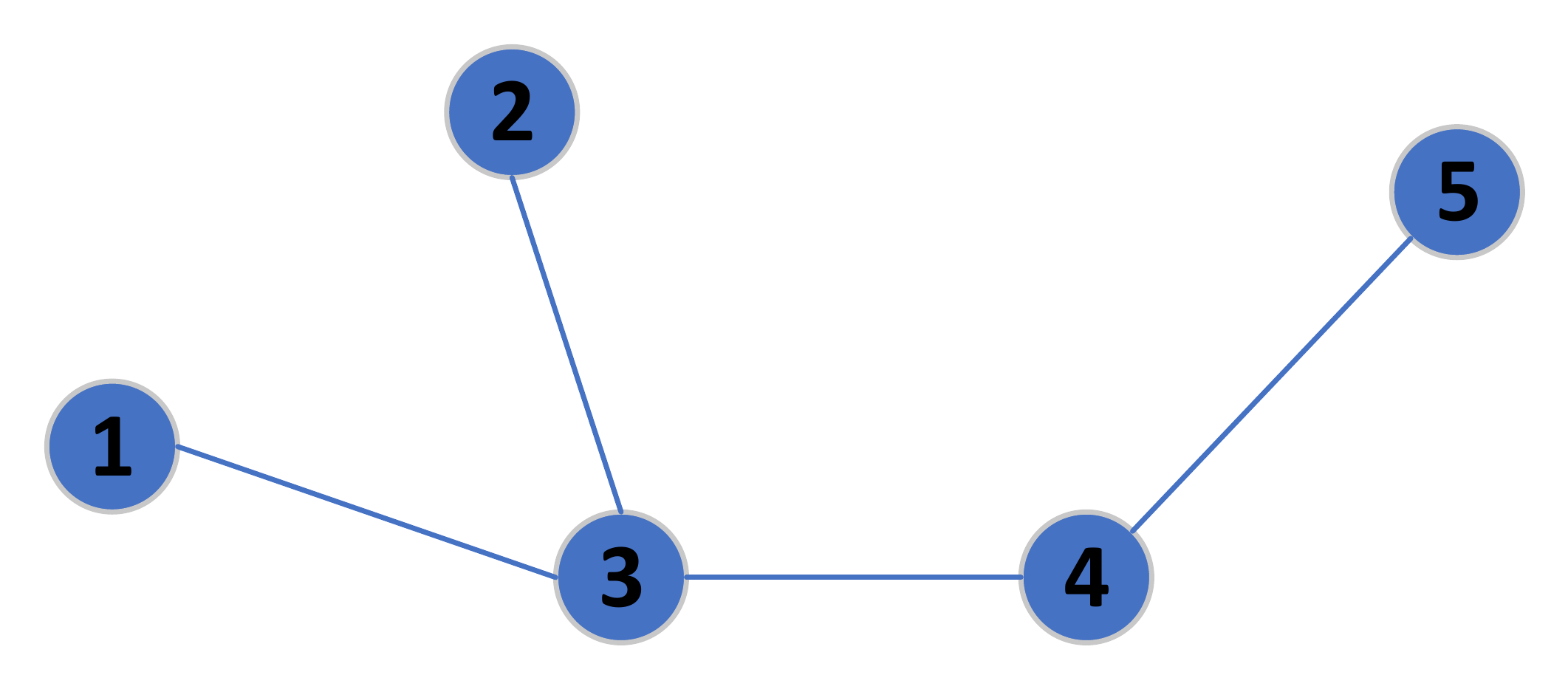}
		\caption{The example of  network topology considered in Sec. \ref{sec:P2P-OAC-SGD and MAC-OAC-SGD Algorithms}.}
		\label{topology}
		\vspace{-0.5cm}
	\end{figure}	
	
	%
	%
 %
	%
	%

	\section{P2P-SGD and MAC-OAC-SGD Algorithms}
	\label{sec:P2P-OAC-SGD and MAC-OAC-SGD Algorithms}

	%
	%
	Instead, we focus on drawing a connection between graph coloring and the duration of the communication block $T$.
	\begin{figure*}
		 \centering
		\begin{subfigure}{0.45\linewidth}
			\centering
			\includegraphics[scale=0.32]{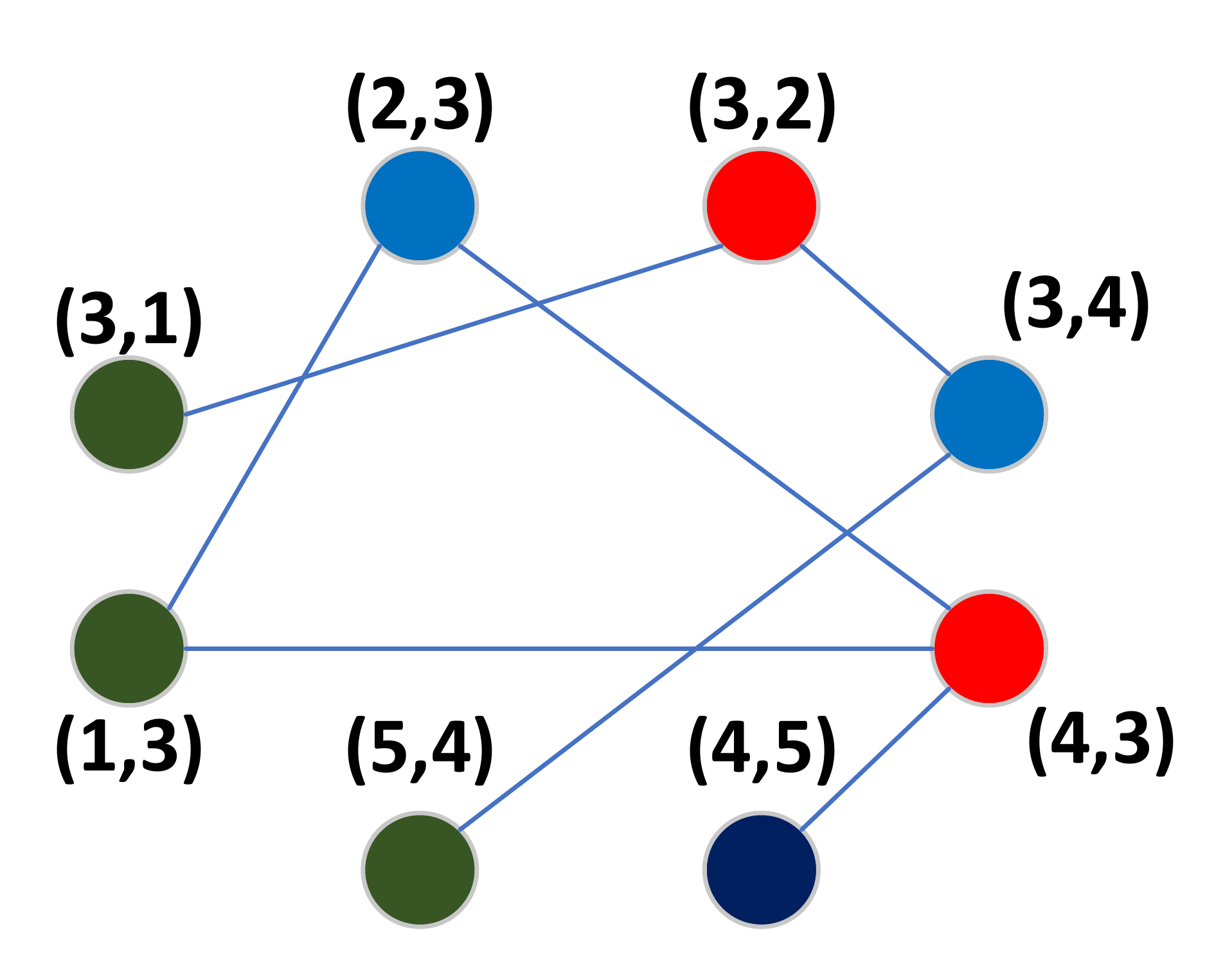}
			\caption{Coloring on $\widetilde{\Gcal}$ for the  P2P-SGD  algorithm in Sec. \ref{sec:P2P-OAC-SGD Algorithm}.}
			\label{colorp2p}
		\end{subfigure}
		\begin{subfigure}{0.5\linewidth}
			\centering
			\includegraphics[scale=0.32]{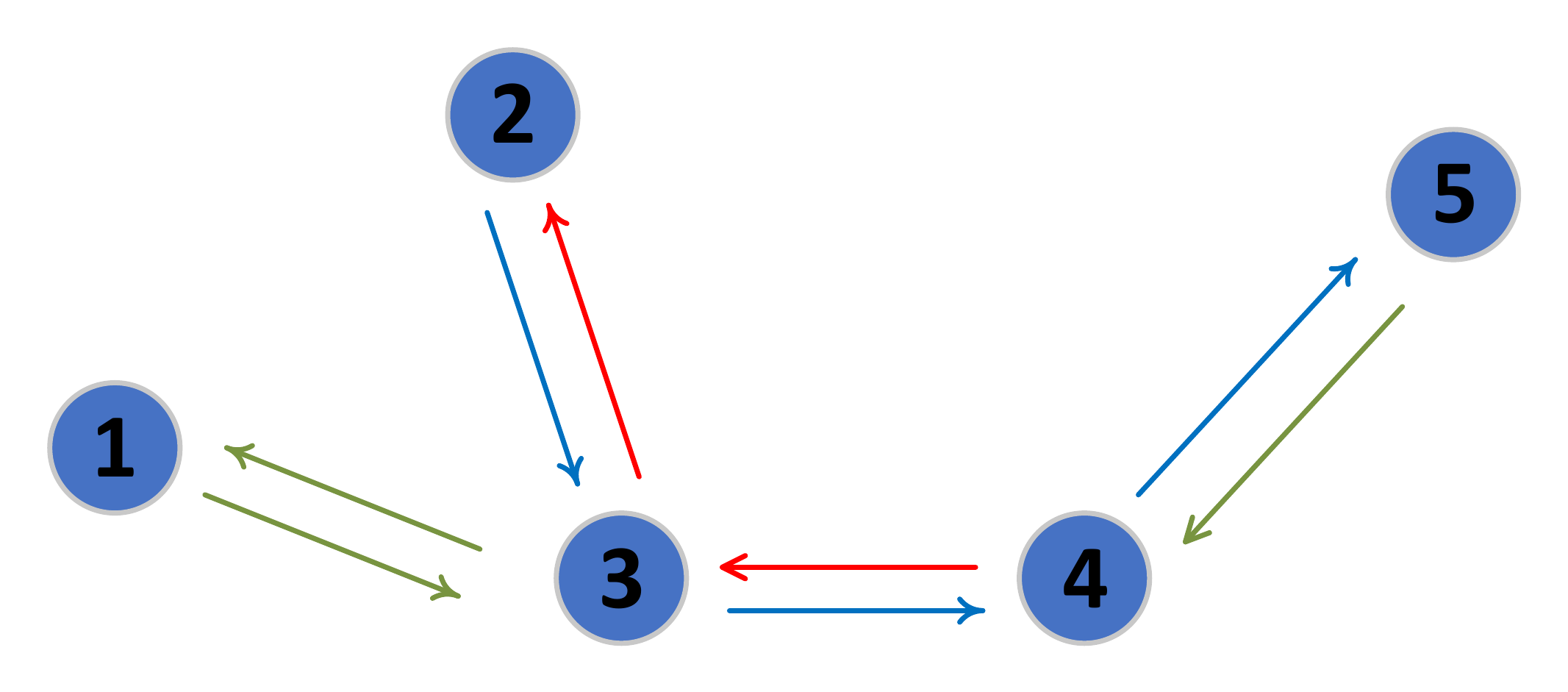}
			\caption{Schedule according to the coloring of $\widetilde{\Gcal}$ in Fig. \ref{colorp2p}.}
			\label{transp2p}
		\end{subfigure}
		\begin{subfigure}{0.5\linewidth}
			\centering
			\includegraphics[scale=0.35]{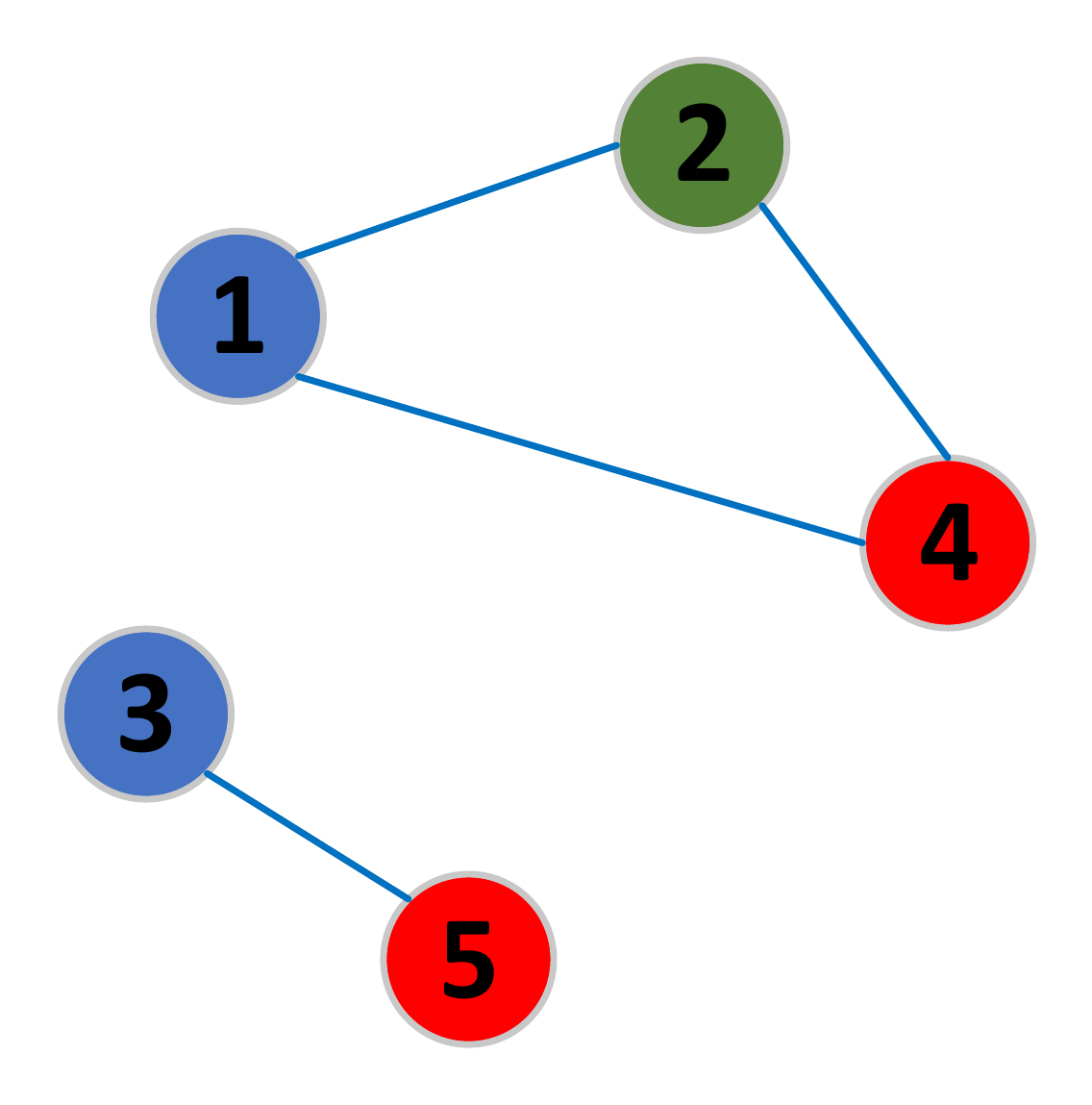}
			\caption{Coloring of the graph $\widehat{\Gcal}$ for the  MAC-OAC-SGD algorthim in Sec. \ref{sec:MAC-OAC-SGD Algorithm}.}
			\label{colormac}
		\end{subfigure}
		\begin{subfigure}{0.45\linewidth}
			\centering
			\includegraphics[scale=0.32]{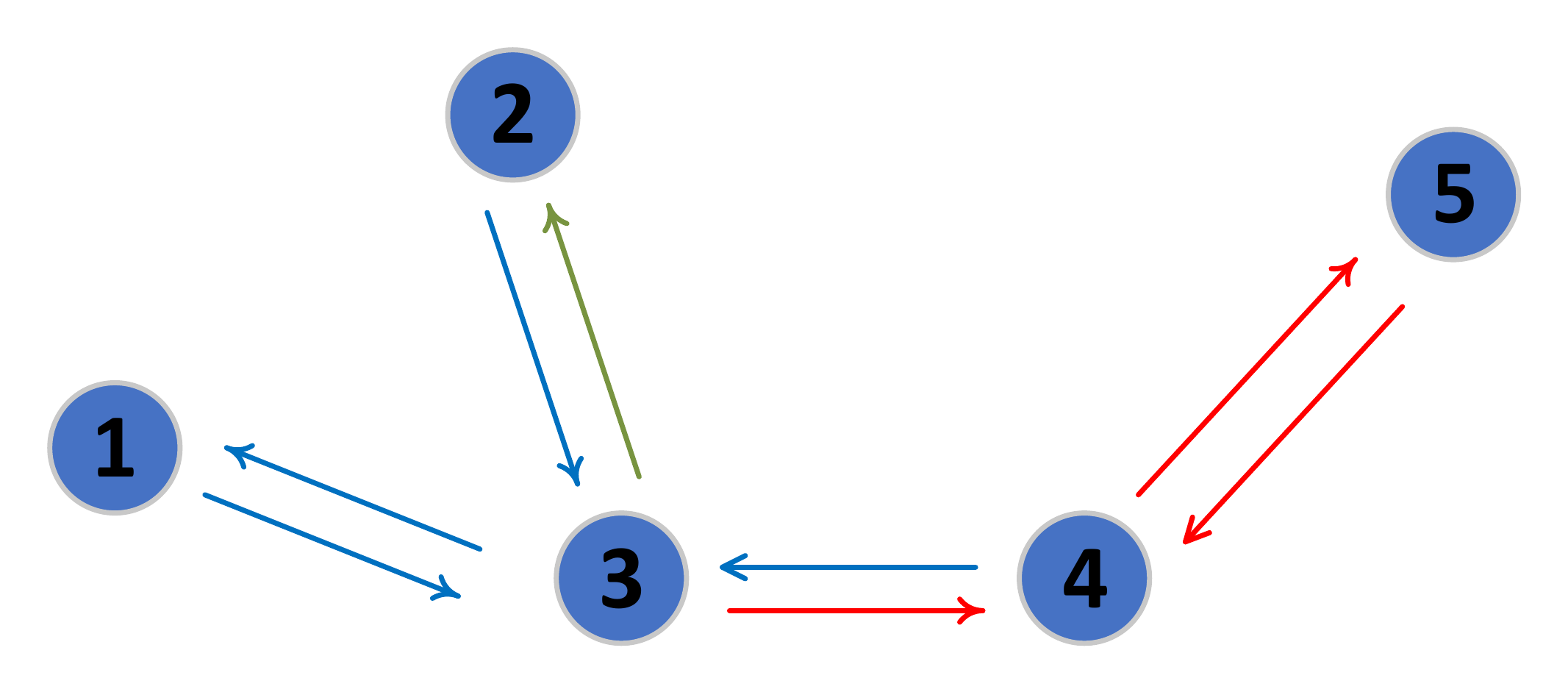}
			\caption{Schedule according to the coloring of $\widehat{\Gcal}$ in Fig. \ref{colormac}.}
			\label{transmac}
		\end{subfigure}
		\caption{An illustration of the schedules for the P2P-SGD and  MAC-OAC-SGD schemes for the example in Sec. \ref{sec:An Illustrative Example}.}
		\label{color}	
	\end{figure*}

%

	\subsection{P2P-SGD Algorithm}
	\label{sec:P2P-OAC-SGD Algorithm}
	For the P2P scheme, transmissions are constrained as follows: in each communication slot (i) the neighborhood of an active receiver node (a.k.a. RX node) contains only one active transmitter node (a.k.a. TX node), and  (ii) the neighborhood of an active TX node contains only one  RX node. Here the neighborhood is defined on graph $\mathcal{G}$.
	We can formulate the scheduling problem as a graph coloring problem as follows: we construct an augmented graph $\widetilde{\Gcal}  \triangleq \Gcal(\Vt,\Et)$ with $\Vt = E$,  where the vertex $(i,j)$ in $\widetilde{\Gcal}$ represents node $i$ transmitting to node $j$ on the initial graph $\Gcal$.
	 Two vertices $(i,j)$, $(l,m) \in \widetilde{\Gcal}$ are connected in $\widetilde{\Gcal}$  if $i=l$ or  $j=m$.
It can be shown	that a solution of the P2P scheduling problem is then obtained by the vertex coloring of $\widetilde{\Gcal}$ by letting $T$ be the number of colors and by having nodes colored with the same color correspond to TX/RX pairs communicating over the same slot.

\subsection{MAC-OAC-SGD Algorithm}
	\label{sec:MAC-OAC-SGD Algorithm}
	 In MAC-OAC-SGD, for each transmission slot two conditions must hold:
	(i) all neighbors  of a RX node must be TX nodes, and (ii) if two nodes are RX nodes, then they cannot have a common neighbor.
	To cast these constraints as a graph theoretical problem, we define the graph  $\widehat{\Gcal} (V,\Eh)$ as the augmented version of the graph $\Gcal(V,\E)$, which have the same vertex set with $\Gcal$ while $\Eh$ is obtained as
	\begin{equation}\label{mac_c1}
	(i,j) \in \Eh\ \text{ if } \  \exists \  k\in V: (i,k),(j,k)\in E.
	\end{equation}
	Again the edge coloring of this graph yields a feasible schedule for the MAC-SDG scheme; this time the nodes colored with the same color correspond to RX nodes active in the same time slot.

	\subsection{An Illustrative Example}
	\label{sec:An Illustrative Example}
	
	To clarify the scheduling strategies for the P2P and MAC schemes, we illustrate them on a simple five-node network in Fig. \ref{topology}, which is obtained by removing the links whose quality is below the threshold $h_{th}$.
	%
	For this network, the graph $\widetilde{\Gcal}$ in Sec. \ref{sec:P2P-OAC-SGD Algorithm} is illustrated in Fig. \ref{colorp2p}. In Fig. \ref{transp2p} we plot the scheduling corresponding to the coloring in Fig. \ref{colorp2p}.
	Note that the TX/RX pairs 
	$(1,3)$, $(3,1)$ and $(5,4)$ have the same color (green) which implies that transmission from nodes $1,3$ and $5$ can be take place is the same time slot. 
	This is represented in Fig. \ref{transp2p} by the green arrows leaving this set of nodes. 
	These transmissions are received at all neighbors, but utilized only at nodes  $3,1$ and $4$, respectively.
%
%
	The same representation is used in Fig. \ref{colormac} and Fig \ref{transmac} for the  MAC-OAC-SGD algorthim.

\addtolength{\topmargin}{.1 in}
	\begin{table*}
\centering
\begin{tabular}{|l|l|l|l|l|l|l|l|l|l|l|l|l|}
\hline
\multirow{3}{*}{} &\multicolumn{6}{l|}{$\sigma=2$}                                                 & \multicolumn{6}{l|}{$\sigma=5$}                                                 \\ \cline{2-13} 
                  & \multicolumn{2}{l|}{$\tau=0.8\sigma$} & \multicolumn{2}{l|}{$\tau=0.12\sigma$} & \multicolumn{2}{l|}{$\tau=1.6\sigma$} & \multicolumn{2}{l|}{$\tau=0.8\sigma$} & \multicolumn{2}{l|}{$\tau=1.2\sigma$} & \multicolumn{2}{l|}{$\tau=1.6\sigma$} \\ \cline{2-13} 
                Scenario &  MAC  & P2P          &     MAC      &   P2P        &   MAC        &    P2P       &   MAC        &   P2P        &    MAC       &    P2P       &   MAC        &    P2P                 \\ \hline
                $\To$  &  20         &  57         &    19       &   28        &  13         &  13          &     20      &  57         &     19      &    28       & 13           &    13       \\ \hline
                Test Accuracy  &    97.48       &  11.21         &    97.48       &   9.8        &   97.51        &   12.8        &    97.47       &    95.26       &       97.52    &   95.28        &    97.48       &   93.50        \\ \hline
\end{tabular}
\caption{Computation and Communication performance of MAC and P2P schemes over 50 trials}
\label{tab:tab1}
\end{table*}

	\section{Numerical Results}
	In our numerical analysis, we consider the image classification problem over the MNIST dataset \cite{mnist} with 10 different image classes, and train a neural network  architecture with two convolutional layers followed by two fully connected layers according to the decentralized SGD framework.  We consider a topology with $n=20$ nodes where  the training dataset is divided among them in an i.i.d. manner. For the training, we use the learning rate $\alpha=0.1$ and measure the test accuracy over 250 iterations.\\
	The network among the nodes is generated as follows: 
	 Matrix $\Mv$ of $n \times n$ i.i.d. Rayleigh random variables with scaling parameter $\sigma \in \{2,5\}$ is first generated. From this, matrix $\Av$ is obtained by setting to zero all the entries of $\Mv$ below threshold $\tau \in \{0.8, 1.2, 1.6\}\sigma$. This truncation induces the probability of an edge as $p \in \{0.7,0.5,0.25\}$.
	The matrix $\Av$ is the matrix of the elements of $\Mv$ above the given threshold. 
	Among the matrix so generated, only the connected matrices $\Av$ are considered. 
	The logic behind this choice of models is as follows: we assume that the $p$ is inversely proportional to $\Ebb[h_{ij}]$ as the threshold $\tau$ indicates the quality of the wireless links used in transmission. 
%
%
%
%

For each sub-scenario, the training/test procedure is repeated $50$ times by generating a new channel realization at the beginning of each time training.
The  performance of the algorithm is measured in terms of the test accuracy and the average value of $T$, i.e. $\To$.
The results of the simulations are presented in Table \ref{tab:tab1}.

Form Table \ref{tab:tab1} we observe that the  MAC-OAC-SGD scheme consistently requires smaller values of $T$ compared to  P2P-SGD. 
	This is expected as in each time slot multiple nodes can communicate.	
	For example, when $\tau=0.8\sigma$, P2P scheme requires almost twice the communication resources as the MAC scheme. 
	On the other, as the connection probability decreases, the advantages of the MAC strategy also diminish. 
	We also observe that, in all the scenarios, the MAC-OAC-SGD scheme achieves a higher average test accuracy compared to the P2P scheme since it is more robust to additive noise.
	This is the case since the P2P accumulates the additive noise over multiple transmission instances, each time the model is exchanged, while the MAC is more robust to additive noise.
	%

\begin{figure}
		\centering
		\includegraphics[scale=0.3]{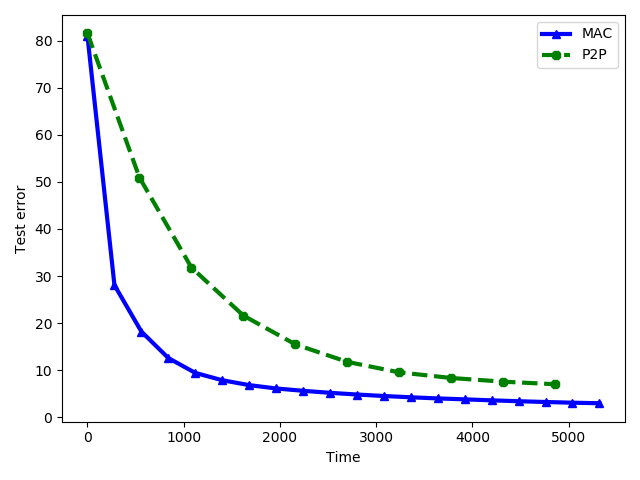}
		\vspace{-0.1 cm}
		\caption{Test error for P2P and MAC schemes for $\sigma=5$ and the random topology based on $\tau=0.8\sigma$.}
		\label{decent_scale4}
				\vspace{-0.35 cm}
	\end{figure}

\begin{figure}
		\centering
		\includegraphics[scale=0.3]{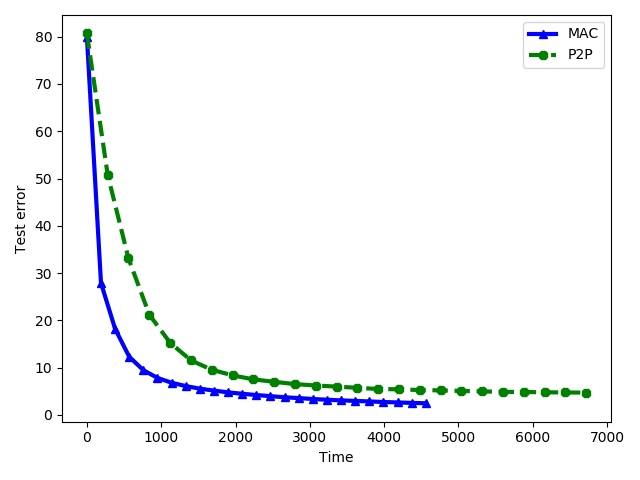}
		\vspace{-0.1 cm}
		\caption{Test error for P2P and MAC schemes for $\sigma=5$ and the random topology based on $\tau=1.2\sigma $.}
		\label{decent_scale6}
				\vspace{-0.5 cm}
	\end{figure}

From the simulation results given Table \ref{tab:tab1}, we can conclude that  advantages of the MAC scheme over the P2P scheme is two-folded; first due to over-the-air aggregation of the signals MAC scheme better suppress the noise, hence can achieve lower test error. Second, compared to P2P scheme, on the average, it requires less number time slots for the communication. 
The joint impact of aforementioned  advantages of MAC scheme are illustrated  in Fig. \ref{decent_scale4} and Fig. \ref{decent_scale6}.
	
	\section{Conclusions}
	In this work, we studied  decentralized SGD  with over-the-air aggregation where the  nodes transmit their local model in a synchronized fashion that all the transmitted models naturally add up  on-the-air at the receiver side. We show that in the scope of decentralized SGD framework implemented in wireless network topology, MAC scheme outperforms the conventional P2P  scheme by reducing the per iteration communication time as well as suppressing the noise term by aggregating the signals on-the-air.

	\bibliographystyle{IEEEtran}
	\bibliography{over_the_air}

\end{document}